\documentclass{article}

\usepackage{microtype}
\usepackage{graphicx}
\usepackage{subcaption}
\usepackage{booktabs}
\usepackage{float}

\usepackage{hyperref}

\usepackage[accepted]{icml2026}

\usepackage{amsmath}
\usepackage{amssymb}
\usepackage{mathtools}
\usepackage{amsthm}
\usepackage{bm}

\usepackage[capitalize,noabbrev]{cleveref}

\theoremstyle{plain}

\theoremstyle{definition}

\theoremstyle{remark}

\icmltitlerunning{Microlensing Detection and Inference via Learned Bayes Factors}

\begin{document}

\twocolumn[
  \icmltitle{Microlensing Detection and Inference via Learned Bayes Factors}

  \begin{icmlauthorlist}
    \icmlauthor{Nolan Smyth}{ciela,mila,udem}
    \icmlauthor{Laurence Perreault-Levasseur}{ciela,mila,udem,flatiron,pi,tsi}
    \icmlauthor{Yashar Hezaveh}{ciela,mila,udem,flatiron,pi,tsi}
  \end{icmlauthorlist}

  \icmlaffiliation{ciela}{Ciela Institute, Montreal Institute for Astrophysics and Machine Learning, Montr\'eal QC, Canada}
  \icmlaffiliation{mila}{Mila - Quebec AI Institute, Montr\'eal QC, Canada}
  \icmlaffiliation{udem}{Department of Physics, Universit\'e de Montr\'eal, Montr\'eal QC, Canada}
  \icmlaffiliation{flatiron}{Center for Computational Astrophysics, Flatiron Institute, New York, USA}
  \icmlaffiliation{pi}{Perimeter Institute for Theoretical Physics, Waterloo, Canada}
  \icmlaffiliation{tsi}{Trottier Space Institute, McGill University, Montr\'eal, Canada}

  \icmlcorrespondingauthor{Nolan Smyth}{nolan.smyth@umontreal.ca}

  \icmlkeywords{Machine Learning, ICML, Simulation-Based Inference, Microlensing, Evidence Networks}

  \vskip 0.3in
]

\printAffiliationsAndNotice{}

\begin{abstract}
We present a unified framework for gravitational microlensing event detection and parameter inference. Traditional pipelines use deterministic hard cuts on photometric statistics, systematically missing low-magnification events in the finite-source regime. We instead frame detection as Bayesian model comparison using Evidence Networks, which learn calibrated Bayes factors from binary-labeled simulations, and combine this with Neural Posterior Estimation (NPE) for amortized parameter inference. Both share a transformer encoder that handles irregularly-sampled time series without imputation. On simulated Roman Space Telescope data, our Evidence Network achieves $99.9\%$ detection efficiency with a false-positive rate below $6\times10^{-4}$ on simulated data with augmentation and noise, but no astrophysical confounders. Gains are most dramatic in the extreme finite-source regime ($\rho \gtrsim 5$), where detection rates reach ${\sim}95\%$ versus ${\sim}65\%$ for hard cuts, precisely the short-duration free-floating planet events most constraining for formation scenarios. Our NPE provides calibrated posteriors, working towards real-time analysis at survey scale.
\end{abstract}

\begin{figure*}[!t]
    \centering
    \includegraphics[width=0.8\linewidth]{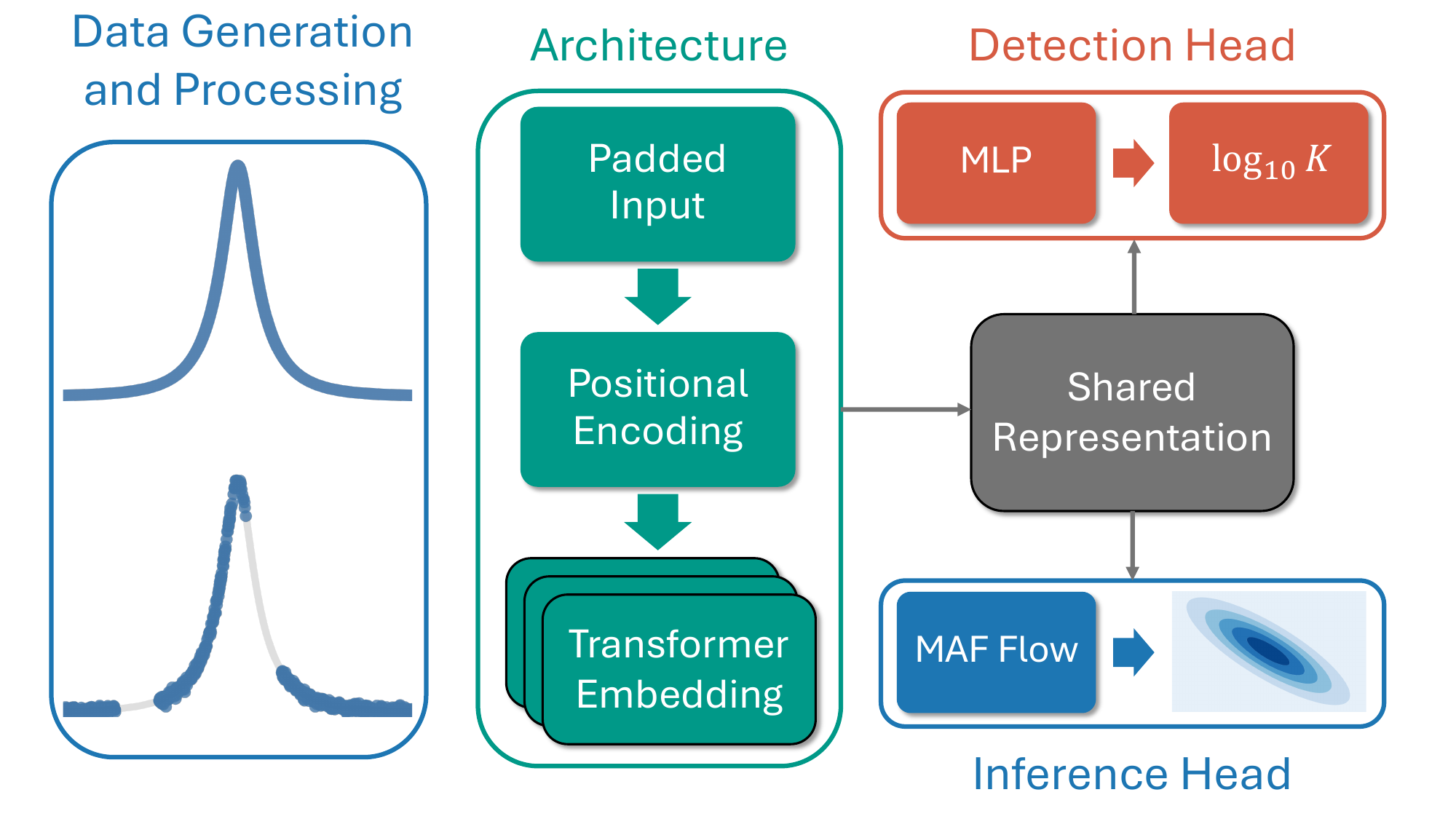}
    \caption{Joint architecture for microlensing detection and parameter inference. Simulated light curves undergo data augmentation and padding. A shared transformer encoder with multi-head self-attention processes the irregularly-sampled time series into a fixed-dimensional summary statistic, the shared representation $\mathbf{s}$. Two specialized heads operate on $\mathbf{s}$. The detection branch outputs calibrated Bayes factors $\log_{10}K$, while the inference branch uses a masked-autoregressive normalizing flow to estimate the full posterior $p(\theta|x)$. This unified framework enables both tasks to benefit from shared features learned by the transformer.}
    \label{fig:sbi}
\end{figure*}

\section{Introduction}
\label{sec:intro}

The Nancy Grace Roman Space Telescope will conduct an unprecedented Galactic Exoplanet Survey targeting over 200 million stars toward the Galactic Bulge \citep{pennyPredictionsWFIRSTMicrolensing2019}. Observing at 15-minute cadence over five 72-day seasons, Roman will revolutionize our understanding of exoplanet demographics through gravitational microlensing, a technique uniquely sensitive to low-mass planets, including terrestrial-mass worlds and free-floating planets (FFPs) unbound to any host star. Yield predictions suggest Roman will detect $\sim$1,400 bound exoplanets \citep{pennyPredictionsWFIRSTMicrolensing2019} and $\sim$250 FFPs \citep{johnsonPredictionsNancyGrace2020}, with lens mass and distance measurements achievable for a significant fraction of detected hosts \citep{terryPredictionsNancyGrace2025a}, enabling direct tests of planet formation and dynamical ejection scenarios \citep{zwartOriginFreefloatingObjects2024,colemanPredictingGalacticPopulation2024}. FFPs are expected to outnumber bound terrestrial planets by a factor of 20 or more \citep{sumiFreeFloatingPlanetMass2023}, and the first direct mass measurement of a free-floating planet via microlensing parallax \citep{dongFreefloatingplanetMicrolensingEvent2026} was recently achieved. Characterizing the low-mass end of the FFP mass function requires sensitivity to short-duration low-magnification events, making performance in this regime crucial to key science drivers.

Realizing this science return requires solving two coupled challenges: \emph{detection} (identifying microlensing signals amid billions of light curves dominated by photometric noise and astrophysical contaminants) and \emph{inference} (estimating model parameters corresponding to those signals). Traditional approaches address these problems sequentially and with fundamentally different methodologies. Detection often relies on deterministic \emph{hard cuts} that apply threshold-based criteria to photometric summary statistics. Representative examples include requiring $\Delta \chi^2 \geq 300$ (difference between microlensing model and flat baseline) combined with $N_{\text{consec}} \geq 6$ consecutive points exceeding $3\sigma$ above baseline \citep{johnsonPredictionsNancyGrace2020}, or variants using the sum of significances over consecutive high-SNR points as a metric (e.g.\ \citealp{sumiUnboundDistantPlanetary2011, mrozNoLargePopulation2017a}). While computationally efficient, hard cuts have limitations. In particular, they are ad-hoc, provide no probabilistic interpretation (a light curve either passes or fails, with no uncertainty quantification), and they can exhibit strong parameter-dependence, missing low-magnification or short-duration events.

For detected events, characterization traditionally employs Markov Chain Monte Carlo (MCMC) sampling to explore the posterior distribution over physical parameters. MCMC provides asymptotically exact Bayesian inference, but requires inference-time computation. This means running MCMCs from scratch for each light curve scales poorly with number of detections. Moreover, MCMC characterization occurs \emph{after} detection, decoupling model comparison (signal vs.\ noise) from parameter estimation in a way that discards information: the Bayes factor for signal detection naturally arises from the same marginal likelihood integrals computed during MCMC.

This work presents a unified neural framework that addresses both challenges within a single probabilistic formalism. We combine \emph{Evidence Networks} \citep{jeffreyEvidenceNetworksSimple2024} for fast Bayes factor estimation with \emph{Neural Posterior Estimation} (NPE) \citep{papamakariosFast$e$freeInference2018, greenbergAutomaticPosteriorTransformation2019} for amortized posterior inference, both sharing a transformer-based encoder for irregularly-sampled time series. Evidence Networks frame detection as supervised Bayesian model comparison: given labeled training data $\{(\mathbf{x}_i, m_i)\}$ with $m_i \in \{0,1\}$ indicating noise ($\mathcal{M}_0$) vs.\ signal ($\mathcal{M}_1$), a network learns the Bayes factor $K \equiv p(\mathbf{x}|\mathcal{M}_1) / p(\mathbf{x}|\mathcal{M}_0)$. Crucially, this requires \emph{only binary labels}, making the approach applicable to scenarios where model evidences are slow or intractable.

Prior work has applied NPE to binary microlensing \citep{zhangRealTimeLikelihoodFreeInference2021}, transformer embeddings for single-lens inference on irregular time-series data \citep{smythTransformerEmbeddingsFast2025}, and normalizing flows with reversible-jump MCMC for microlensing model selection \citep{keehanMicrolensingModelInference2022}. Our framework extends these by unifying detection and characterization within a single amortized pipeline (Section~\ref{sec:architecture}).

Our contributions are threefold: (i) a unified probabilistic framework for detection and characterization that enables targeted scientific questions with calibrated uncertainty (e.g.\ the probability a light curve contains an FFP with $t_E < 1$ day); (ii) the first application of Evidence Networks to irregularly-sampled astrophysical time series, recovering calibrated Bayes factors from binary labels alone and replacing ad hoc $\Delta\chi^2$ thresholds; and (iii) on simulated Roman-like data, improved detection completeness in the extreme finite-source regime ($\rho \gtrsim 5$) relative to literature hard cuts (Section~\ref{sec:microlensing}).\footnote{The code used in this work is available at: \url{https://github.com/NolanSmyth/evidence_networks_sbi_microlensing}.}

\section{Methods}
\label{sec:methods}

Our framework consists of two complementary neural networks operating on the same input representation: an Evidence Network for signal detection via Bayes factor estimation, and a Neural Posterior Estimator for inference.

\subsection{Data and Modeling}

We model Finite-Source Point-Lens (FSPL) events using \texttt{ESPLMag2} as implemented in \texttt{VBMicrolensing} \citep{bozzaVBBinaryLensingPublicPackage2018a}, which uses a pre-computed table, the entries of which are calculated via contour integration. Note that this is also the backend behind \texttt{FSPLargemodel} in \texttt{pyLIMA}, another microlensing modeling package \citep{bacheletPyLIMAOpenSource2017}.

The five microlensing parameters we consider are $\boldsymbol{\theta} = (t_0, u_0, \log_{10} t_E, \log_{10} \rho, f_s)$: the time of closest lens-source alignment (days), the impact parameter in units of the Einstein radius $\theta_E$, the Einstein crossing time (days), the normalized source radius $\rho \equiv \theta_s / \theta_E$ (with $\theta_s$ the angular source size), and the source flux fraction $f_s \in [0, 1]$. The observed flux at time $t$, normalized to the baseline, is

\begin{equation}
F(t) = f_s A(t, \boldsymbol{\theta}) + (1 - f_s)
\end{equation}

\noindent where $A(t, \boldsymbol{\theta})$ is the magnification factor computed accounting for finite size of the source, and $1-f_s$ represents blended light from unresolved stars common in crowded fields.

For simplicity, we adopt box-uniform priors: $t_0 \in [0, 20]$ days, $u_0 \in [0, 1.5]$, $\log_{10}(t_E) \in [-1, 1.3]$ (0.1--20 days, spanning short-duration FFP candidate events), $\log_{10}(\rho) \in [-2, 1]$ (source radii from $0.01\theta_E$ to $10\theta_E$), and $f_s \in [0.1, 1.0]$. We stress that for a production-ready pipeline, the priors should be informed by galactic and population models. For example, if FFPs follow a power-law mass function \citep{sumiFreeFloatingPlanetMass2023}, one would expect a much higher density of short-duration events whose duration is determined by the angular size of the source, rather than a peaked mass function \citep{deroccoReconstructingFreefloatingPlanet2025} which would predict fewer events with $\rho \gg 1$. Ideally, these factors should also be incorporated into a self-consistent hierarchical model, rather than being chosen ad hoc.

Each light curve is represented as a padded sequence of length $L = 1000$ with three channels per timestep: $\mathbf{x}_i = (t_i^{\text{norm}}, F_i, \sigma_i)$ for $i = 1, \ldots, L$. Times are normalized to $[-1, 1]$ over the 20-day window via $t^{\text{norm}} = 2t/T_{\text{total}} - 1$, allowing the network to learn position-invariant features. $F_i$ is the normalized flux relative to the baseline (taken to be unity), and $\sigma_i$ is the per-point photometric uncertainty. Padded positions take a value of $-2$ in all channels, enabling the transformer to identify valid data via attention masking.

\subsection{Data Augmentation}
\label{sec:augmentation}

To demonstrate performance on some realistic observing conditions, we employ on-the-fly data augmentation during training. For each parameter sample $\boldsymbol{\theta}$ drawn from the prior:

\begin{enumerate}
    \item \textbf{Base simulation}: Generate a dense, noiseless light curve using a cadence of 15 minutes (1920 points uniformly spaced over 20 days)
    \item \textbf{Seasonal gaps}: Introduce 0--3 gaps of length 1--10 days each, simulating seasonal observability windows or downtime
    \item \textbf{Random dropout}: Remove a random fraction (0--60\%) of remaining points, mimicking bad readings or cadence variations
    \item \textbf{Photometric noise}: Add Gaussian noise with per-point standard deviation $\sigma$ drawn uniformly from $[0.001, 0.02]$ in relative flux units (corresponding to $\sim$0.1--2\%, covering the typical expected photometric precision of Roman)
\end{enumerate}

We apply recoverability filters to exclude training samples with insufficient coverage. Events must satisfy: at least 5 points within $1.5 t_E$ of $t_0$ (peak region coverage), at least 5 points beyond $3.0 t_E$ from $t_0$ (baseline establishment), and \textit{peak} signal-to-noise ratio SNR $> 5$, where SNR is computed as $(A_{\text{peak}} - 1) / \sigma_{\text{baseline}}$. These criteria ensure the network trains on physically recoverable events while learning to handle realistic data quality.

\subsection{Transformer Embedding Network}
\label{sec:architecture}

The key challenge is converting variable-length, irregularly-sampled light curves into fixed-dimensional summaries that preserve the temporal structure needed for both detection and parameter inference. Both the Evidence Network and Neural Posterior Estimator share a common embedding architecture based on the transformer encoder \citep{vaswaniAttentionAllYou2023}. This design offers several advantages. First, self-attention learns long-range dependencies: a point at $t = 5$ days can directly attend to baseline measurements at $t = 15$ days, capturing the full event morphology. Second, attention weights are permutation-invariant but position-aware through encodings; the network learns what matters about the temporal structure, not fixed grid positions. Third, the shared embedding serves both detection and inference, enabling transfer learning: features useful for computing Bayes factors also help constrain parameters.

The network processes 3-channel input $\mathbf{X} \in \mathbb{R}^{B \times L \times 3}$ (batch size $B$, sequence length $L$) to produce a summary $\mathbf{s} \in \mathbb{R}^{B \times d_{\text{model}}}$. Each timestep $\mathbf{x}_i = (t_i^{\text{norm}}, F_i, \sigma_i)$ is first projected from 3 dimensions to $d_{\text{model}} = 256$ via a learned linear layer $\mathbf{h}_i^{(0)} = \mathbf{W}_{\text{proj}} \mathbf{x}_i + \mathbf{b}_{\text{proj}}$, scaled by $\sqrt{d_{\text{model}}}$ for variance stabilization. This projection creates a high-dimensional representation where the network can learn complex relationships between flux, time, and uncertainty.

We append a learnable classification token $\mathbf{c} \in \mathbb{R}^{d_{\text{model}}}$ to the sequence. This token serves as an information aggregator: through self-attention, it attends to all timesteps and accumulates evidence for detection or parameter constraints. Its final hidden state becomes our summary $\mathbf{s}$.

We add sinusoidal positional encodings \citep{vaswaniAttentionAllYou2023}; our normalized time channel already provides explicit positional cues, so we empirically observed that these encodings offer only marginal benefit, but speed up convergence somewhat. The core architecture is 6 transformer encoder layers with 8-head self-attention and position-wise feedforward networks ($d_{\text{ff}} = 512$), with standard residual connections and layer normalization. Padded positions (marked by time $= -2$) are masked in the attention computation, so the network handles variable-length sequences without interpolation or imputation. We extract the CLS token's final hidden state and apply layer normalization.

\subsection{Evidence Networks for Signal Detection}

\subsubsection{Bayesian Framework}

Unlike $\Delta\chi^2$ statistics, which have no rigorous Bayesian interpretation for the non-nested model comparison, Bayes factors have a clear meaning. We treat microlensing event detection as a model selection problem: given observed data $\mathbf{x}$ (a light curve), we compare the signal hypothesis $\mathcal{M}_1$ (microlensing event present) against the noise-only hypothesis $\mathcal{M}_0$. The posterior probability of the event model relative to the noise-only model is

\begin{equation}
\frac{p(\mathcal{M}_1|\mathbf{x})}{p(\mathcal{M}_0 |\mathbf{x})} =\frac{p(\mathbf{x}|\mathcal{M}_1)}{p(\mathbf{x}|\mathcal{M}_0)}\frac{p(\mathcal{M}_{1})}{p(\mathcal{M}_{0})} = K \frac{p(\mathcal{M}_{1})}{p(\mathcal{M}_{0})},
\end{equation}
where $p(\mathcal{M}_1)/p(\mathcal{M}_0)$ is the prior probability of any given light curve containing a true event.

The Bayes factor $K$ quantifies the relative evidence, or how much to update your model prior given the data:

\begin{equation}
K = \frac{p(\mathbf{x}|\mathcal{M}_1)}{p(\mathbf{x}|\mathcal{M}_0)} = \frac{\int p(\mathbf{x}|\boldsymbol{\theta}, \mathcal{M}_1) p(\boldsymbol{\theta}|\mathcal{M}_1) \, d\boldsymbol{\theta}}{p(\mathbf{x}|\mathcal{M}_0)}
\end{equation}

\noindent where the numerator marginalizes over the prior $p(\boldsymbol{\theta}|\mathcal{M}_1)$ to compute the model evidence. Unlike $\Delta\chi^2$ evaluated at maximum-likelihood parameters, this marginalization automatically penalizes model complexity via the Occam factor: a larger parameter space must spread its predictive probability over more possibilities, reducing the evidence unless the data strongly favor the signal model. Computing this integral analytically or via sampling is often slow or intractable in practice. However, an Evidence Network trained on binary labeled training data alone can learn to quickly and accurately predict $\log K$ directly from the data.

\subsubsection{Architecture and Training}

We train the Evidence Network using the leaky parity-odd power (l-POP) exponential loss with exponent $\alpha = 2$, following the recommendations of \citet{jeffreyEvidenceNetworksSimple2024}:

\begin{equation}
\mathcal{L}(f(\mathbf{x}), m) = \exp\left[\left(\frac{1}{2} - m\right) J_\alpha(f(\mathbf{x}))\right]
\end{equation}

\noindent where $m \in \{0, 1\}$ is the binary label (1 for signal, 0 for noise), $f(\mathbf{x})$ is the network output, and $J_\alpha(z) = z + z|z|^{\alpha-1}$ is the l-POP transform. For the optimal network $f^*$ minimizing the expected loss, we have:

\begin{equation}
\log K = J_\alpha(f^*(\mathbf{x})).
\end{equation}

The l-POP loss is not the only choice that leads to an estimate of the Bayes factor, but it does spread the dynamic range of Bayes factors across many orders of magnitude, preventing saturation for strong signals while maintaining sensitivity to weaker signals.

The Evidence Network itself consists of the transformer embedding followed by a 3-layer multilayer perceptron with hidden dimensions $[256, 128, 1]$ and ReLU activations. The final layer outputs a scalar $f(\mathbf{x})$ transformed to $\log_{10} K$ via l-POP.

We train for 200 epochs using the Adam optimizer \citep{kingmaAdamMethodStochastic2017} with learning rate $10^{-4}$, batch size 1024, and early stopping based on validation loss (patience 20 epochs). Training data consists of 100,000 simulated light curves with equal signal/noise fractions, generated on-the-fly with augmentation as described in Section~\ref{sec:augmentation}.

The network is trained using \emph{only binary labels}, not the true Bayes factors. In real microlensing applications, ground-truth model evidences are unavailable; labels typically come from human classification, conventional detection algorithms, or simulations. The analytic $\log K$ values in our toy model (Appendix~\ref{sec:toy}) serve solely for post-hoc validation. This constraint ensures the method is applicable to realistic scenarios where Bayes factors cannot be computed.

\begin{figure*}[!t]
    \centering
    \includegraphics[width=0.95\linewidth]{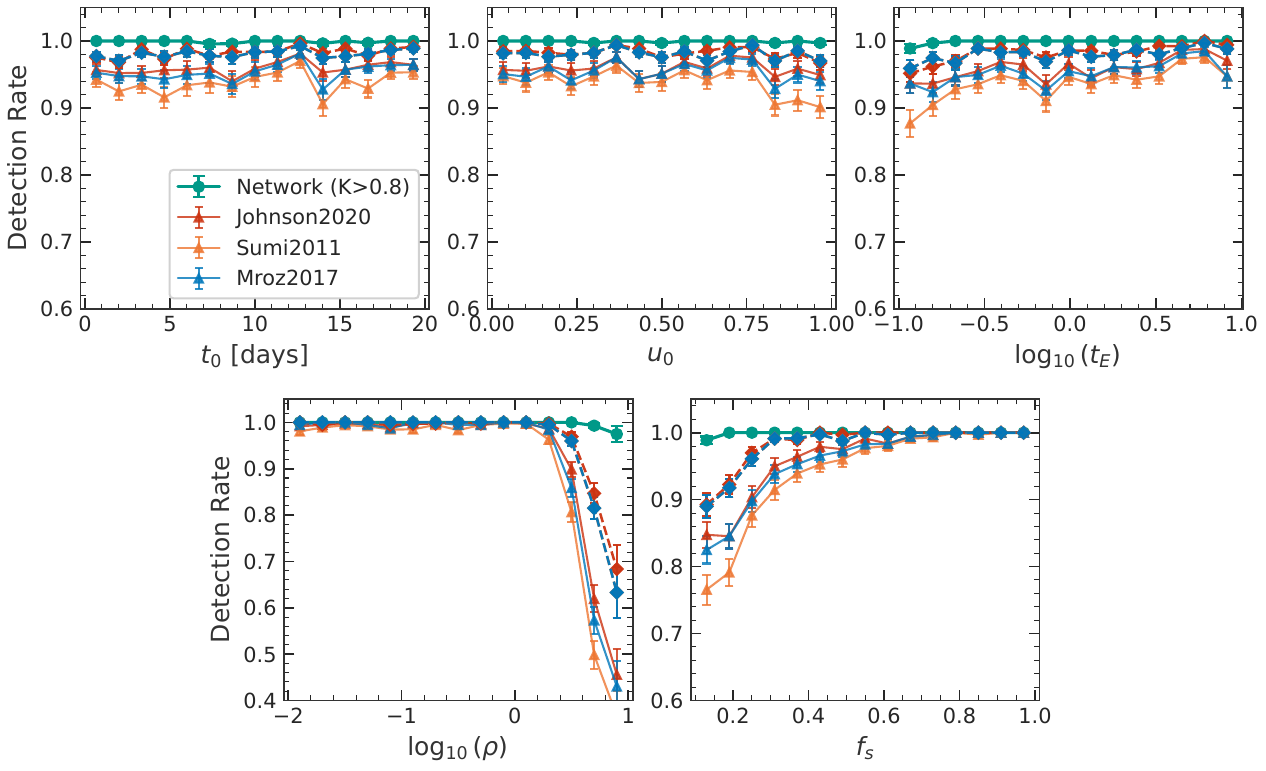}
    \caption{Detection efficiency as a function of FSPL parameters for the Evidence Network and literature hard cuts. The Evidence Network maintains high efficiency in the extreme finite-source regime ($\rho \gtrsim 5$) where hard cuts systematically fail. Both as-published and re-tuned hard cuts are shown with re-tuned shown as dashed lines; hard-cut statistics use oracle (true) parameters, so the comparison is conservative. Error bars denote per-bin binomial $68\%$ confidence intervals.}
    \label{fig:detection_rate}
\end{figure*}

\subsection{Neural Posterior Estimation for Parameter Inference}
\label{sec:npe}

For detected events, we estimate the posterior $p(\boldsymbol{\theta}|\mathbf{x})$ using simulation-based inference. We employ Neural Posterior Estimation, which directly learns the posterior via conditional density estimation rather than explicitly modeling the likelihood. The posterior estimator combines the transformer embedding network with a Masked Autoregressive Flow (MAF) normalizing flow \citep{papamakariosMaskedAutoregressiveFlow2018a}. The transformer produces a summary statistic $\mathbf{s} = g_{\phi}(\mathbf{x}) \in \mathbb{R}^{256}$, which conditions the flow, so that $p_\psi(\boldsymbol{\theta}|\mathbf{x}) \approx p_\psi(\boldsymbol{\theta}|\mathbf{s})$, with
\begin{equation}
p_\psi(\boldsymbol{\theta}|\mathbf{s}) = p_{\mathbf{z}}\!\left(T_\psi^{-1}(\boldsymbol{\theta}|\mathbf{s})\right) \left|\det \frac{\partial T_\psi^{-1}}{\partial \boldsymbol{\theta}}\right|,
\end{equation}

\noindent where $T_\psi: \mathbb{R}^5 \times \mathbb{R}^{256} \to \mathbb{R}^5$ is the learned invertible transformation (flow), $p_{\mathbf{z}} = \mathcal{N}(\mathbf{0}, \mathbf{I})$ is a standard Gaussian base distribution, and the Jacobian determinant accounts for the change of variables. The MAF architecture uses 10 autoregressive layers, each a masked affine coupling layer with context embedding $\mathbf{s}$ injected at every layer.

The autoregressive property ensures efficient density evaluation: the $i$-th parameter dimension is generated conditioned on previous dimensions $\theta_{1:i-1}$ and the summary $\mathbf{s}$, enabling tractable likelihood computation via:

\begin{equation}
\log p_\psi(\boldsymbol{\theta}|\mathbf{s}) = \log p_{\mathbf{z}}(T_\psi^{-1}(\boldsymbol{\theta}|\mathbf{s})) + \log \left|\det \frac{\partial T_\psi^{-1}}{\partial \boldsymbol{\theta}}\right|
\end{equation}

\noindent Intuitively, the log-posterior at $\boldsymbol{\theta}$ is the log-density of the base Gaussian evaluated at the point that maps to $\boldsymbol{\theta}$ under the flow, plus a Jacobian correction for the change of variables.

We train via maximum likelihood on 80,000 simulated pairs $\{(\boldsymbol{\theta}_i, \mathbf{x}_i)\}_{i=1}^{80000}$, with an additional 20,000 held out for validation, using the \texttt{sbi} library \citep{boeltsSbiReloadedToolkit2025}. We jointly optimize the transformer parameters $\phi$ and flow parameters $\psi$ so that simulated parameter-data pairs receive high posterior density under the model:

\begin{equation}
\max_{\phi, \psi} \, \mathbb{E}_{(\boldsymbol{\theta}, \mathbf{x}) \sim p(\boldsymbol{\theta}, \mathbf{x})} \left[ \log p_\psi(\boldsymbol{\theta}|g_\phi(\mathbf{x})) \right]
\end{equation}

Training uses the same data augmentation pipeline (Section~\ref{sec:augmentation}), ensuring the network learns to handle gaps, dropout, and variable noise. We employ the Adam optimizer with learning rate $10^{-4}$, batch size 512, and train for up to 400 epochs with early stopping (patience 30 epochs) based on validation negative log-likelihood. Training requires approximately 18 hours on a single NVIDIA H100 GPU.

At test time, drawing posterior samples for a new observation requires only a single forward pass through the transformer to compute $\mathbf{s} = g_\phi(\mathbf{x})$, followed by sampling from the flow $\boldsymbol{\theta}^{(i)} \sim p_\psi(\cdot|\mathbf{s})$ by transforming base samples $\mathbf{z}^{(i)} \sim \mathcal{N}(\mathbf{0}, \mathbf{I})$ via $\boldsymbol{\theta}^{(i)} = T_\psi(\mathbf{z}^{(i)}|\mathbf{s})$. Generating 10,000 posterior samples takes $\sim21$~ms per event, a $\sim$$16{,}000\times$ speedup over a comparably-tuned \texttt{emcee} sampler on the same hardware (NVIDIA H100); see Appendix~\ref{app:npe_timing} for details.

\section{Application to Microlensing Detection and Characterization}
\label{sec:microlensing}

Before applying the method to microlensing, we validate it on a toy Gaussian-bump detection problem where the Bayes factor admits a closed-form solution. The Evidence Network recovers near-oracle ROC performance (AUC = 0.725 vs.\ 0.726) and produces well-calibrated posteriors, confirming that binary labels alone suffice to recover calibrated Bayes factor estimates; full details and figures are in Appendix~\ref{sec:toy}. Having validated the methodology, we apply the complete framework to simulated microlensing light curves. This section presents results for both signal detection (via Bayes factors) and parameter characterization (via SBI posteriors).

\subsection{Detection Performance vs.\ Literature Hard Cuts}

Traditional microlensing surveys employ deterministic detection criteria that apply threshold-based cuts to photometric summary statistics. As noted in Section~\ref{sec:intro}, these ``hard cuts'' lack probabilistic interpretation and exhibit strong parameter-dependence that systematically misses certain event types. To quantify these limitations, we compare our Evidence Network against methods from the literature.

The first method, used in \citet{johnsonPredictionsNancyGrace2020} for the Nancy Grace Roman Space Telescope yield predictions, requires both $\Delta\chi^2 \geq 300$ (where $\Delta\chi^2$ measures the improvement in fit between a microlensing model and a flat baseline) and $N_{\rm consec} \geq 6$ consecutive photometric points exceeding $3\sigma$ above baseline. The second criterion, used in \citet{sumiUnboundDistantPlanetary2011} for the MOA-II Galactic Bulge survey, requires $\chi_{3+} \geq 80$ (the sum of significances over all consecutive high-SNR points) combined with $N_{\rm consec} \geq 3$. Finally, \citet{mrozNoLargePopulation2017a} adopted a more permissive threshold of $\chi_{3+} \geq 32$ with $N_{\rm consec} \geq 3$ for the OGLE survey.

For a fair comparison, we compute these statistics on the same test set of 5,000 simulated events with data augmentations as described in Section~\ref{sec:augmentation}. Critically, when evaluating the hard cuts, we calculate $\Delta\chi^2$ using the \emph{true} injected FSPL parameters; this is an oracle upper bound, since in practice one must fit the model, incurring optimization failures and local minima that would further degrade detection rates. We apply the literature thresholds as published: the \citet{johnsonPredictionsNancyGrace2020} cuts are Roman-tuned at the cadence level ($N_{\rm consec}=6$ at 15-minute sampling) but were calibrated against Roman's full multi-season observing strategy rather than a 20-day window, while the \citet{sumiUnboundDistantPlanetary2011, mrozNoLargePopulation2017a} thresholds inherit from ground-based surveys. To ensure the comparison does not merely penalize thresholds calibrated for a different observing setup, we additionally \emph{re-tune} each hard cut (jointly sweeping the statistic thresholds) on a validation set drawn from the same generator used to train the network (Appendix~\ref{app:ablation}, Table~\ref{tab:tuned_cuts}). The Evidence Network threshold is likewise selected on validation data.

For the Evidence Network, we adopt a detection threshold of $\log_{10} K > 0.8$ (equivalently, $K > 6.3$, or approximately 1 in a million chance of misclassification under ideal conditions). This threshold was selected to achieve zero false positives on our validation set, ensuring high purity for candidate detections; on the held-out test set the network likewise yields no false positives among 5,000 Gaussian-noise events, corresponding to a $95\%$ upper limit on the false-positive rate of $6\times10^{-4}$. With this conservative criterion, the Evidence Network detects \textbf{4,996 of 5,000 signals (99.9\%)}, outperforming all hard cuts: \citet{johnsonPredictionsNancyGrace2020} achieves 95.9\%, \citet{sumiUnboundDistantPlanetary2011} 93.9\%, and \citet{mrozNoLargePopulation2017a} 95.4\%. Even after re-tuning, the evidence network significantly outperforms the hard cuts, despite the cuts retaining the advantage of using oracle (true) parameters.

The network's advantage is most pronounced in the extreme finite-source regime ($\rho \gtrsim 5$), where detection efficiency remains $\sim 95\%$ compared to $\sim 65\%$ for hard cuts. This regime dominates short-duration events, precisely the population of greatest interest for constraining the FFP mass function. Hard cuts fail here because finite-source events have lower peak magnification despite distinctive morphological signatures; the Evidence Network extracts evidence from the full temporal structure rather than relying on peak flux alone.

Figure~\ref{fig:detection_rate} breaks this down by FSPL parameter: the Evidence Network maintains consistently high detection rates across the parameter space, while hard cuts degrade significantly in the finite-source regime. The hard cuts also show modest degradation for highly-blended events (low $f_s$), another important case for Roman's densely populated fields of view toward the Galactic Bulge. Representative events detected only by the network (Appendix~\ref{app:network_only}) illustrate the failure modes: low peak magnification, observation gaps interrupting consecutive high-SNR points, and ongoing events without an established baseline.

\subsection{Posterior Estimation: Injected vs.\ Recovered Parameters}

\begin{figure*}[!t]
    \centering
    \includegraphics[width=0.95\linewidth]{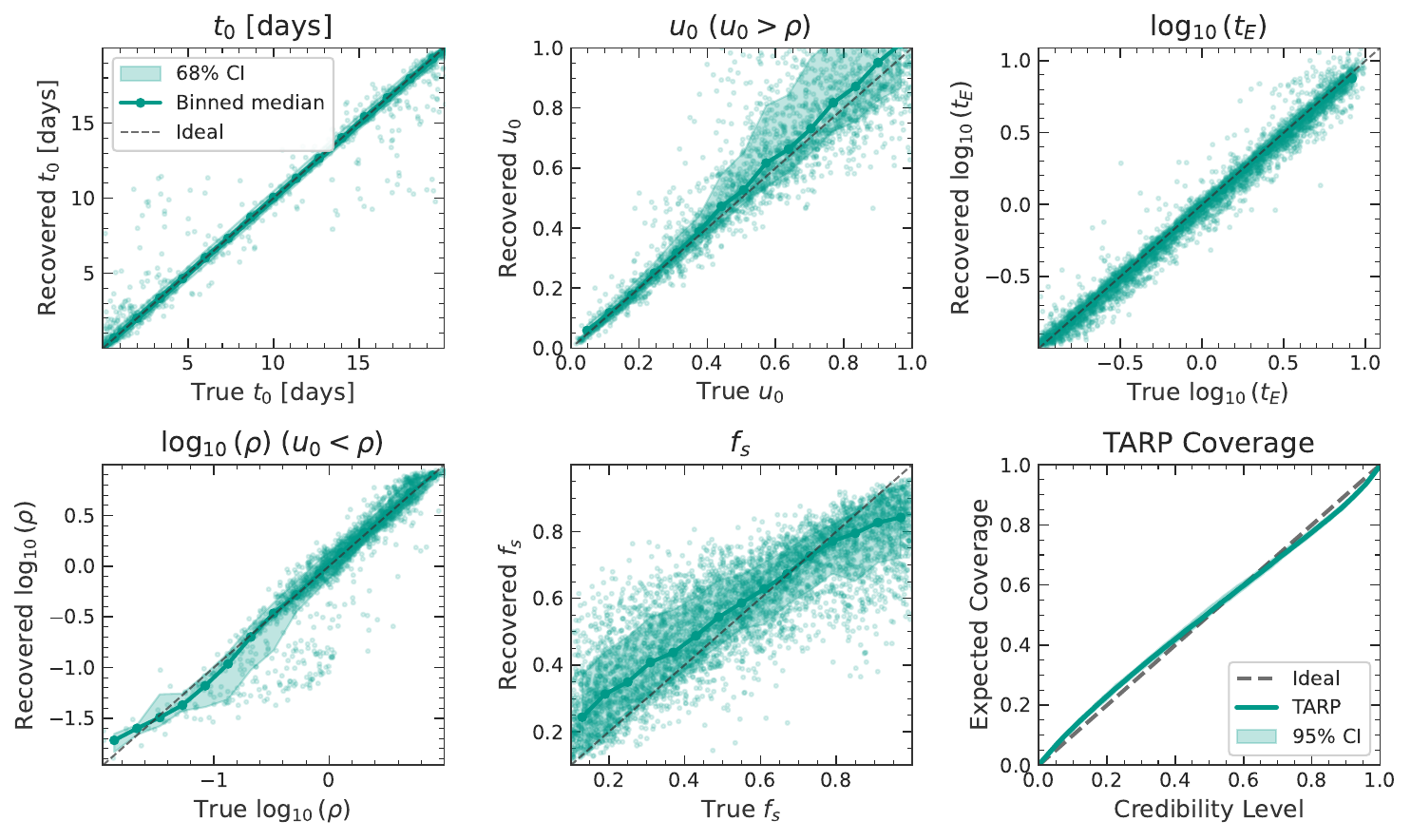}
    \caption{Posterior calibration assessment for the events detected by the Evidence Network. \textbf{Top row and middle-left:} Injected vs. recovered parameter values show strong correlation with ideal 1:1 line (gray dashed). Shaded regions show 68\% credible intervals. Note physical degeneracies: $u_0$ is only constrained when $u_0 > \rho$ (impact parameter exceeds source size), while $\rho$ is only constrained when $u_0 < \rho$ (finite-source effects dominate). \textbf{Bottom right:} TARP coverage test demonstrates good calibration and coverage.}
    \label{fig:ivr_tarp}
\end{figure*}

For events classified as likely signals by the Evidence Network, we estimate posterior distributions $p(\bm{\theta}|\bm{x})$ over the five physical parameters using the transformer-based Neural Posterior Estimator (Section~\ref{sec:npe}). The key question for amortized inference is whether the network produces posteriors that are both \emph{accurate} (centered on the true parameter values) and \emph{calibrated} (credible intervals reflect genuine uncertainty rather than overconfident or underconfident predictions). We assess both properties on the 4,996 events detected by the Evidence Network at our chosen threshold ($\log_{10} K > 0.8$).

Figure~\ref{fig:ivr_tarp} shows injected versus recovered parameter values for all detected events. The correlation between injected and median recovered values is high for all five parameters. Binned median trends (shaded regions) closely track the identity line (dashed gray), with only mild deviations at the edges.

\textbf{Physical degeneracies.} When the source is much smaller than the impact parameter ($\rho \ll u_0$), finite-source effects are negligible and $\rho$ is poorly constrained. Conversely, when $u_0 < \rho$ (source larger than impact parameter), the event is dominated by finite-source effects and $u_0$ becomes degenerate. We therefore only plot a subset of events for these two parameters corresponding to the non-degenerate regions. The blend parameter shows the largest scatter as expected: blending primarily affects the amplitude normalization. Nonetheless, the posteriors correctly identify highly-blended vs.\ minimally-blended events. We quantify the residual biases and degeneracy axes in Appendix~\ref{app:bias}.

\textbf{Calibration and Coverage}
To rigorously validate posterior calibration, we apply the Test of Accuracy with Random Points (TARP) \citep{lemosSamplingBasedAccuracyTesting2023}. The bottom right panel of Figure~\ref{fig:ivr_tarp} shows that the expected coverage closely tracks the credibility level of the posteriors, demonstrating that they are not biased and are reasonably well-calibrated.

\section{Discussion}
\label{sec:discussion}

Our unified framework replaces ad hoc hard cuts with calibrated Bayes factors via Evidence Networks, combined with amortized posterior inference via NPE.

\subsection{Advantages of the Unified Framework}

Our approach combines signal detection and parameter characterization within a single probabilistic framework, offering several advantages over traditional microlensing analysis pipelines:

\textbf{Transferable representations.} The two heads share an encoder but are trained separately, so the framework is ``unified'' at the level of the learned representation rather than a single joint objective. We show in Appendix~\ref{app:ablation} that this sharing is beneficial, not merely architectural: an Evidence head trained on the frozen, SBI-trained encoder transfers from inference to detection at essentially no cost, matching the headline detection rate at a fraction of the training budget and wall-clock cost.

\textbf{Amortized inference.} Both the Evidence Network and SBI posterior estimator require upfront computational investment for training (18--20 hours total on a single GPU). However, once trained, inference for new observations is nearly instantaneous: detecting signals via Bayes factors requires $\sim$10 milliseconds per light curve, while drawing 10,000 posterior samples requires $\sim$100 milliseconds. This one-time training cost amortizes over billions of light curves, enabling real-time analysis at survey scale.

\textbf{Natural handling of irregular data.} The transformer architecture processes light curves as unordered sets of observations with attention masking, requiring no interpolation to fixed grids or imputation of missing data. This is particularly valuable for microlensing surveys, which may exhibit seasonal gaps, variable cadence driven by observing strategy, and detector-specific systematics.

\textbf{Principled model comparison.} The Evidence Network outputs calibrated model posterior probabilities $p(\mathcal{M}_1|\mathbf{x})$, enabling principled decision-making under uncertainty: observers can set detection thresholds based on desired false positive rates or incorporate prior odds to account for event rarity, naturally trading off completeness against purity.

\subsection{Current Limitations and Future Extensions}

\textbf{False positives and non-Gaussian noise.} Our training data contains only microlensing signals and Gaussian noise, so the framework does not currently model astrophysical false positives such as stellar flares, eclipsing binaries, cataclysmic variables, pulsating variables, or instrumental systematics. The reported false-positive rate is therefore an upper bound for the signal-detection problem in isolation, not a survey-level false-positive rate, which in practice contains contaminants that are challenging to distinguish from true positives. For instance, \citet{mrozTESSFreefloatingPlanet2024} argued on prior grounds that TESS event TIC-107150013 is a stellar flare from a magnetically active giant, despite the light curve strongly favoring a finite-source microlensing fit over flare models ($\Delta\mathrm{BIC}\gtrsim 13$) \citep{kunimotoSearchingFreeFloatingPlanets2025}; resolving such tensions demands combining calibrated likelihoods with realistic astrophysical priors, not light curve vetting alone. Real survey data also exhibit correlated noise, cosmic rays, and bad pixels. Both limitations are in principle addressable: contaminants can be injected into training, and the noise distribution can be learned from real observations \citep{leginGaussianNoiseGeneralized2023, leginGravitationalWaveParameterEstimation2024}. More generally, SBI is sensitive to distributional shifts \citep{filippRobustnessNeuralRatio2024}, so production priors should be informed by galactic population models rather than the box-uniform priors used here.

\textbf{Fixed observing window.} The network is trained on 20-day windows with $t_E \sim 0.1$--20 days, limiting applicability to longer-duration events ($t_E \sim 30$--200 days) from more massive lenses. Extension requires variable-length windows with adaptive padding or sliding-window tiling of long light curves.

\textbf{Binary lens models.} Binary lenses produce complex caustic morphologies; \citet{zhangRealTimeLikelihoodFreeInference2021} demonstrated NPE for binary parameter estimation but not detection. Extending our framework requires higher-dimensional parameter spaces and potentially hierarchical model comparison.

\textbf{Population-level inference.} Our framework performs event-level inference, whereas scientific questions (e.g., the FFP mass function) require hierarchical models $p(\boldsymbol{\phi}|\{\mathbf{x}_i\})$. Recent hierarchical SBI work \citep{ericksonLensModelingSTRIDES2024, perkinsDisentanglingBlackHole2024} points toward natural extensions.

\subsection{Comparison with Prior Work}

\textbf{Evidence Networks.} \citet{jeffreyEvidenceNetworksSimple2024} introduced Evidence Networks for cosmological model comparison using regularly-sampled CMB power spectra and galaxy survey statistics. Our work represents the first application to irregularly-sampled astrophysical time series, demonstrating that the l-POP-Exponential loss generalizes to variable-length sequential data when combined with transformer embeddings. The key architectural innovation is replacing fixed-length dense networks with attention-based aggregation over unordered observations.

\textbf{SBI for microlensing.} \citet{zhangRealTimeLikelihoodFreeInference2021} applied NPE to binary microlensing parameter estimation but did not address signal detection or single-lens events. Our contribution is twofold: (1) extending to the full single-lens FSPL parameter space with rigorous TARP validation, and (2) unifying detection and characterization within a single framework via Evidence Networks.

\textbf{Differentiable microlensing.} Recent work has introduced differentiable microlensing simulators that enable gradient-based inference. \texttt{microJAX} \citep{miyazakiMicroJAXDifferentiableFramework2025} provides a fully differentiable JAX-based ray-shooting implementation, while \texttt{microlux} \citep{renDifferentiableBinaryMicrolensing2025} implements differentiable contour integration for binary lenses, with \citet{renKMT2025BLG1314KMT2025BLG1392Two2026} demonstrating its application to real events via Hamiltonian Monte Carlo. These approaches are complementary to ours: differentiable simulators enable gradient-based samplers but require a differentiable forward model, while our SBI framework operates with any black-box simulator and amortizes inference across all observations.

\textbf{Transformer time-series models.} \citet{salinasDistinguishingPlanetaryTransit2023} employed transformers to classify \textit{TESS} transits vs.\ false positives using attention maps to identify diagnostic features. Our approach differs in using transformers as summary statistic extractors for downstream detection and density estimation, enabling calibrated probabilistic inference with explicit uncertainty quantification.

\subsection{Broader Implications}

The combination of Evidence Networks and transformer-based SBI addresses a general challenge in time-domain astronomy: automated, scalable, and principled analysis of irregularly-sampled transient signals. Beyond microlensing, this framework may be applicable to domains such as supernova classification, exoplanet transit detection, variable star labeling, and more. The key requirement is simply a forward simulator capable of generating sufficient training data.

\section*{Impact Statement}
This paper presents work whose goal is to advance the field of Machine Learning. There are many potential societal consequences of our work, none of which we feel must be specifically highlighted here.

\bibliographystyle{icml2026}
\bibliography{references}

@misc{kunimotoSearchingFreeFloatingPlanets2025,
  title = {Searching for {{Free-Floating Planets}} with {{TESS}}: {{Results}} from {{Sectors}} 61-65},
  shorttitle = {Searching for {{Free-Floating Planets}} with {{TESS}}},
  author = {Kunimoto, Michelle and DeRocco, William and Smyth, Nolan and Bryson, Steve and Gaudi, B. Scott},
  year = 2025,
  month = oct,
  number = {arXiv:2404.11666},
  eprint = {2404.11666},
  primaryclass = {astro-ph},
  publisher = {arXiv},
  doi = {10.48550/arXiv.2404.11666},
  urldate = {2026-04-19},
  archiveprefix = {arXiv}
}

@article{dongFreefloatingplanetMicrolensingEvent2026,
  title = {A Free-Floating-Planet Microlensing Event Caused by a {{Saturn-mass}} Object},
  author = {Dong, Subo and Wu, Zexuan and Ryu, Yoon-Hyun and Udalski, Andrzej and Mroz, Przemek and Rybicki, Krzysztof A. and Hodgkin, Simon T. and Wyrzykowski, Lukasz and Eyer, Laurent and Bensby, Thomas and Chen, Ping and Wang, Sharon X. and Gould, Andrew and Yang, Hongjing and Albrow, Michael D. and Chung, Sun-Ju and Han, Cheongho and Hwang, Kyu-Ha and Jung, Youn Kil and Shin, In-Gu and Shvartzvald, Yossi and Yee, Jennifer C. and Zang, Weicheng and Kim, Dong-Jin and Lee, Chung-Uk and Park, Byeong-Gon and Poleski, Radoslaw and Skowron, Jan and Szymanski, Michal K. and Soszynski, Igor and Pietrukowicz, Pawel and Kozlowski, Szymon and Skowron, Dorota M. and Ulaczyk, Krzysztof and Gromadzki, Mariusz and Ratajczak, Milena and Iwanek, Patryk and Wrona, Marcin and Mroz, Mateusz J. and Rixon, Guy and Harrison, Diana L. and Breedt, Elme},
  year = 2026,
  month = jan,
  journal = {Science},
  volume = {391},
  number = {6780},
  eprint = {2601.00057},
  primaryclass = {astro-ph},
  pages = {96--99},
  issn = {0036-8075, 1095-9203},
  doi = {10.1126/science.adv9266},
  urldate = {2026-04-09},
  archiveprefix = {arXiv}
}

@article{keehanMicrolensingModelInference2022,
  title = {Microlensing Model Inference with Normalising Flows and Reversible Jump {{MCMC}}},
  author = {Keehan, D. and Yarndley, J. and Rattenbury, N.},
  year = 2022,
  month = oct,
  journal = {Astronomy and Computing},
  volume = {41},
  pages = {100657},
  issn = {22131337},
  doi = {10.1016/j.ascom.2022.100657},
  urldate = {2026-04-09},
  langid = {english}
}

@misc{miyazakiMicroJAXDifferentiableFramework2025,
  title = {{{microJAX}}: {{A Differentiable Framework}} for {{Microlensing Modeling}} with {{GPU-Accelerated Image-Centered Ray Shooting}}},
  shorttitle = {{{microJAX}}},
  author = {Miyazaki, Shota and Kawahara, Hajime},
  year = 2025,
  month = oct,
  number = {arXiv:2510.02639},
  eprint = {2510.02639},
  primaryclass = {astro-ph},
  publisher = {arXiv},
  doi = {10.48550/arXiv.2510.02639},
  urldate = {2026-04-09},
  archiveprefix = {arXiv}
}

@article{renDifferentiableBinaryMicrolensing2025,
  title = {A Differentiable Binary Microlensing Model Using Adaptive Contour Integration Method},
  author = {Ren, Haibin and Zhu, Wei},
  year = 2025,
  month = mar,
  journal = {The Astronomical Journal},
  volume = {169},
  number = {3},
  eprint = {2501.07268},
  primaryclass = {astro-ph},
  pages = {170},
  issn = {0004-6256, 1538-3881},
  doi = {10.3847/1538-3881/adb1b2},
  urldate = {2026-04-09},
  archiveprefix = {arXiv}
}

@misc{renKMT2025BLG1314KMT2025BLG1392Two2026,
  title = {{{KMT-2025-BLG-1314}} and {{KMT-2025-BLG-1392}}: Two Microlensing Planetary/Brown-Dwarf Candidates Analyzed with Differentiable Code},
  shorttitle = {{{KMT-2025-BLG-1314}} and {{KMT-2025-BLG-1392}}},
  author = {Ren, Haibin and Zang, Weicheng and Zhu, Wei and Ryu, Yoon-Hyun and Tang, Yuchen and Zhang, Jiyuan and Albrow, Michael D. and Chung, Sun-Ju and Gould, Andrew and Han, Cheongho and Hwang, Kyu-Ha and Jung, Youn Kil and Shin, In-Gu and Shvartzvald, Yossi and Yang, Hongjing and Yee, Jennifer C. and Kim, Dong-Jin and Lee, Chung-Uk and Park, Byeong-Gon and Tang, Yunyi and Maoz, Dan and Mao, Shude and Qian, Qiyue},
  year = 2026,
  month = mar,
  number = {arXiv:2603.01735},
  eprint = {2603.01735},
  primaryclass = {astro-ph},
  publisher = {arXiv},
  doi = {10.48550/arXiv.2603.01735},
  urldate = {2026-04-09},
  archiveprefix = {arXiv}
}

@misc{terryPredictionsNancyGrace2025a,
  title = {Predictions of the {{Nancy Grace Roman Space Telescope Galactic Exoplanet Survey}}. {{IV}}. {{Lens Mass}} and {{Distance Measurements}}},
  author = {Terry, Sean K. and Bachelet, Etienne and Zohrabi, Farzaneh and Verma, Himanshu and Crisp, Alison and Huston, Macy and McGee, Carissma and Penny, Matthew and Abrams, Natasha S. and Albrow, Michael D. and Anderson, Jay and Bagheri, Fatemeh and Beaulieu, Jean-Phillipe and Bellini, Andrea and Bennett, David P. and Bergsten, Galen and Bhadra, T. Dex and Bhattacharya, Aparna and Bond, Ian A. and Bozza, Valerio and Brandon, Christopher and Novati, Sebastiano Calchi and Carey, Sean and Christiansen, Jessie and DeRocco, William and Gaudi, B. Scott and Hulberg, Jon and Silva, Stela Ishitani and Jones, Sinclaire E. and Kerins, Eamonn and Khakpash, Somayeh and Kruszynska, Katarzyna and Lam, Casey and Lu, Jessica R. and Malpas, Amber and Miyazaki, Shota and Mroz, Przemek and Murlidhar, Arjun and Nataf, David and Newman, Marz and Olmschenk, Greg and Poleski, Rakek and Ranc, Clement and Rattenbury, Nicholas J. and Rybicki, Krzysztof and Saggese, Vito and Sobeck, Jennifer and Stassun, Keivan G. and Stephan, Alexander P. and Street, Rachel A. and Sumi, Takahiro and Suzuki, Daisuke and Vandorou, Aikaterini and Vyas, Meet and Yee, Jennifer C. and Zang, Weicheng and Zhang, Keming},
  year = 2025,
  month = oct,
  number = {arXiv:2510.13974},
  eprint = {2510.13974},
  primaryclass = {astro-ph},
  publisher = {arXiv},
  doi = {10.48550/arXiv.2510.13974},
  urldate = {2026-04-09},
  archiveprefix = {arXiv}
}

@article{salinasDistinguishingPlanetaryTransit2023,
  title = {Distinguishing a Planetary Transit from False Positives: A {{Transformer-based}} Classification for Planetary Transit Signals},
  shorttitle = {Distinguishing a Planetary Transit from False Positives},
  author = {Salinas, Helem and Pichara, Karim and Brahm, Rafael and {P{\'e}rez-Galarce}, Francisco and Mery, Domingo},
  year = 2023,
  month = jul,
  journal = {Monthly Notices of the Royal Astronomical Society},
  volume = {522},
  number = {3},
  pages = {3201--3216},
  issn = {0035-8711},
  doi = {10.1093/mnras/stad1173},
  urldate = {2025-02-14}
}

@article{leginGaussianNoiseGeneralized2023,
  title = {Beyond {{Gaussian Noise}}: {{A Generalized Approach}} to {{Likelihood Analysis}} with Non-{{Gaussian Noise}}},
  shorttitle = {Beyond {{Gaussian Noise}}},
  author = {Legin, Ronan and Adam, Alexandre and Hezaveh, Yashar and Levasseur, Laurence Perreault},
  year = 2023,
  month = jun,
  journal = {The Astrophysical Journal Letters},
  volume = {949},
  number = {2},
  eprint = {2302.03046},
  primaryclass = {astro-ph},
  pages = {L41},
  issn = {2041-8205, 2041-8213},
  doi = {10.3847/2041-8213/acd645},
  urldate = {2024-09-18},
  archiveprefix = {arXiv}
}

@misc{leginGravitationalWaveParameterEstimation2024,
  title = {Gravitational-{{Wave Parameter Estimation}} in Non-{{Gaussian}} Noise Using {{Score-Based Likelihood Characterization}}},
  author = {Legin, Ronan and Isi, Maximiliano and Wong, Kaze W. K. and Hezaveh, Yashar and {Perreault-Levasseur}, Laurence},
  year = 2024,
  month = oct,
  number = {arXiv:2410.19956},
  eprint = {2410.19956},
  publisher = {arXiv},
  doi = {10.48550/arXiv.2410.19956},
  urldate = {2024-11-18},
  archiveprefix = {arXiv}
}

@misc{papamakariosMaskedAutoregressiveFlow2018a,
  title = {Masked {{Autoregressive Flow}} for {{Density Estimation}}},
  author = {Papamakarios, George and Pavlakou, Theo and Murray, Iain},
  year = 2018,
  month = jun,
  number = {arXiv:1705.07057},
  eprint = {1705.07057},
  primaryclass = {stat},
  publisher = {arXiv},
  doi = {10.48550/arXiv.1705.07057},
  urldate = {2026-01-08},
  archiveprefix = {arXiv}
}

@misc{bacheletPyLIMAOpenSource2017,
  title = {{{pyLIMA}} : An Open Source Package for Microlensing Modeling. {{I}}. Presentation of the Software and Analysis on Single Lens Models},
  shorttitle = {{{pyLIMA}}},
  author = {Bachelet, E. and Norbury, M. and Bozza, V. and Street, R.},
  year = 2017,
  month = sep,
  number = {arXiv:1709.08704},
  eprint = {1709.08704},
  primaryclass = {astro-ph},
  publisher = {arXiv},
  doi = {10.48550/arXiv.1709.08704},
  urldate = {2026-01-08},
  archiveprefix = {arXiv}
}

@misc{smythTransformerEmbeddingsFast2025,
  title = {Transformer {{Embeddings}} for {{Fast Microlensing Inference}}},
  author = {Smyth, Nolan and {Perreault-Levasseur}, Laurence and Hezaveh, Yashar},
  year = 2025,
  month = dec,
  number = {arXiv:2512.11687},
  eprint = {2512.11687},
  primaryclass = {astro-ph},
  publisher = {arXiv},
  doi = {10.48550/arXiv.2512.11687},
  urldate = {2026-01-08},
  archiveprefix = {arXiv}
}

@misc{papamakariosFast$e$freeInference2018,
  title = {Fast \${$\varepsilon\$$}-Free {{Inference}} of {{Simulation Models}} with {{Bayesian Conditional Density Estimation}}},
  author = {Papamakarios, George and Murray, Iain},
  year = 2018,
  month = apr,
  number = {arXiv:1605.06376},
  eprint = {1605.06376},
  primaryclass = {stat},
  publisher = {arXiv},
  doi = {10.48550/arXiv.1605.06376},
  urldate = {2026-01-08},
  archiveprefix = {arXiv}
}

@misc{jeffreyEvidenceNetworksSimple2024,
  title = {Evidence {{Networks}}: Simple Losses for Fast, Amortized, Neural {{Bayesian}} Model Comparison},
  shorttitle = {Evidence {{Networks}}},
  author = {Jeffrey, Niall and Wandelt, Benjamin D.},
  year = 2024,
  month = jan,
  number = {arXiv:2305.11241},
  eprint = {2305.11241},
  publisher = {arXiv},
  doi = {10.48550/arXiv.2305.11241},
  urldate = {2024-12-03},
  archiveprefix = {arXiv}
}

@article{sumiUnboundDistantPlanetary2011,
  title = {Unbound or {{Distant Planetary Mass Population Detected}} by {{Gravitational Microlensing}}},
  author = {Sumi, T. and Kamiya, K. and Udalski, A. and Bennett, D. P. and Bond, I. A. and Abe, F. and Botzler, C. S. and Fukui, A. and Furusawa, K. and Hearnshaw, J. B. and Itow, Y. and Kilmartin, P. M. and Korpela, A. and Lin, W. and Ling, C. H. and Masuda, K. and Matsubara, Y. and Miyake, N. and Motomura, M. and Muraki, Y. and Nagaya, M. and Nakamura, S. and Ohnishi, K. and Okumura, T. and Perrott, Y. C. and Rattenbury, N. and Saito, To and Sako, T. and Sullivan, D. J. and Sweatman, W. L. and Tristram, P. J. and Yock, P. C. M. and Szymanski, M. K. and Kubiak, M. and Pietrzynski, G. and Poleski, R. and Soszynski, I. and Wyrzykowski, L. and Ulaczyk, K.},
  year = 2011,
  month = may,
  journal = {Nature},
  volume = {473},
  number = {7347},
  eprint = {1105.3544},
  primaryclass = {astro-ph},
  pages = {349--352},
  issn = {0028-0836, 1476-4687},
  doi = {10.1038/nature10092},
  urldate = {2026-01-08},
  archiveprefix = {arXiv}
}

@article{mrozNoLargePopulation2017a,
  title = {No Large Population of Unbound or Wide-Orbit {{Jupiter-mass}} Planets},
  author = {Mroz, Przemek and Udalski, Andrzej and Skowron, Jan and Poleski, Radoslaw and Kozlowski, Szymon and Szymanski, Michal K. and Soszynski, Igor and Wyrzykowski, Lukasz and Pietrukowicz, Pawel and Ulaczyk, Krzysztof and Skowron, Dorota and Pawlak, Michal},
  year = 2017,
  month = aug,
  journal = {Nature},
  volume = {548},
  number = {7666},
  eprint = {1707.07634},
  primaryclass = {astro-ph},
  pages = {183--186},
  issn = {0028-0836, 1476-4687},
  doi = {10.1038/nature23276},
  urldate = {2026-01-08},
  archiveprefix = {arXiv}
}

@misc{boeltsSbiReloadedToolkit2025,
  title = {Sbi Reloaded: A Toolkit for Simulation-Based Inference Workflows},
  shorttitle = {Sbi Reloaded},
  author = {Boelts, Jan and Deistler, Michael and Gloeckler, Manuel and {Tejero-Cantero}, {\'A}lvaro and Lueckmann, Jan-Matthis and Moss, Guy and Steinbach, Peter and Moreau, Thomas and Muratore, Fabio and Linhart, Julia and Durkan, Conor and Vetter, Julius and Miller, Benjamin Kurt and Herold, Maternus and Ziaeemehr, Abolfazl and Pals, Matthijs and Gruner, Theo and Bischoff, Sebastian and Krouglova, Nastya and Gao, Richard and Lappalainen, Janne K. and Mucs{\'a}nyi, B{\'a}lint and Pei, Felix and Schulz, Auguste and Stefanidi, Zinovia and Rodrigues, Pedro and Schr{\"o}der, Cornelius and Zaid, Faried Abu and Beck, Jonas and Kapoor, Jaivardhan and Greenberg, David S. and Gon{\c c}alves, Pedro J. and Macke, Jakob H.},
  year = {2025},
  month = aug,
  number = {arXiv:2411.17337},
  eprint = {2411.17337},
  primaryclass = {cs},
  publisher = {arXiv},
  doi = {10.48550/arXiv.2411.17337},
  urldate = {2025-08-24},
  archiveprefix = {arXiv}
}

@article{bozzaVBBinaryLensingPublicPackage2018a,
  title = {{{VBBinaryLensing}}: A Public Package for Microlensing Light Curve Computation},
  shorttitle = {{{VBBinaryLensing}}},
  author = {Bozza, V. and Bachelet, E. and Bartoli{\'c}, F. and Heintz, T. and Hoag, A. and Hundertmark, M.},
  year = {2018},
  month = oct,
  journal = {Monthly Notices of the Royal Astronomical Society},
  volume = {479},
  number = {4},
  eprint = {1805.05653},
  primaryclass = {astro-ph},
  pages = {5157--5167},
  issn = {0035-8711, 1365-2966},
  doi = {10.1093/mnras/sty1791},
  urldate = {2025-08-24},
  archiveprefix = {arXiv}
}

@misc{colemanPredictingGalacticPopulation2024,
  title = {Predicting the {{Galactic}} Population of Free-Floating Planets from Realistic Initial Conditions},
  author = {Coleman, Gavin A. L. and DeRocco, William},
  year = {2024},
  month = jul,
  number = {arXiv:2407.05992},
  eprint = {2407.05992},
  primaryclass = {astro-ph},
  publisher = {arXiv},
  doi = {10.48550/arXiv.2407.05992},
  urldate = {2024-09-30},
  archiveprefix = {arXiv}
}

@misc{deroccoReconstructingFreefloatingPlanet2025,
  title = {Reconstructing the {{Free-floating Planet Mass Function}} with the {{Nancy Grace Roman Space Telescope}}},
  author = {DeRocco, William and Penny, Matthew T. and Johnson, Samson A. and McGill, Peter},
  year = {2025},
  month = apr,
  number = {arXiv:2505.00092},
  eprint = {2505.00092},
  primaryclass = {astro-ph},
  publisher = {arXiv},
  doi = {10.48550/arXiv.2505.00092},
  urldate = {2025-05-09},
  archiveprefix = {arXiv}
}

@misc{ericksonLensModelingSTRIDES2024,
  title = {Lens {{Modeling}} of {{STRIDES Strongly Lensed Quasars}} Using {{Neural Posterior Estimation}}},
  author = {Erickson, Sydney and {Wagner-Carena}, Sebastian and Marshall, Phil and Millon, Martin and Birrer, Simon and Roodman, Aaron and Schmidt, Thomas and Treu, Tommaso and Schuldt, Stefan and Shajib, Anowar and Venkatraman, Padma and Collaboration, The LSST Dark Energy Science},
  year = {2024},
  month = oct,
  number = {arXiv:2410.10123},
  eprint = {2410.10123},
  publisher = {arXiv},
  doi = {10.48550/arXiv.2410.10123},
  urldate = {2024-10-22},
  archiveprefix = {arXiv}
}

@misc{filippRobustnessNeuralRatio2024,
  title = {Robustness of {{Neural Ratio}} and {{Posterior Estimators}} to {{Distributional Shifts}} for {{Population-Level Dark Matter Analysis}} in {{Strong Gravitational Lensing}}},
  author = {Filipp, Andreas and Hezaveh, Yashar and {Perreault-Levasseur}, Laurence},
  year = {2024},
  month = nov,
  number = {arXiv:2411.05905},
  eprint = {2411.05905},
  primaryclass = {astro-ph},
  publisher = {arXiv},
  doi = {10.48550/arXiv.2411.05905},
  urldate = {2025-08-24},
  archiveprefix = {arXiv}
}

@article{foreman-mackeyEmceeMCMCHammer2013,
  title = {Emcee: {{The MCMC Hammer}}},
  shorttitle = {Emcee},
  author = {{Foreman-Mackey}, Daniel and Hogg, David W. and Lang, Dustin and Goodman, Jonathan},
  year = {2013},
  month = mar,
  journal = {Publications of the Astronomical Society of the Pacific},
  volume = {125},
  number = {925},
  eprint = {1202.3665},
  primaryclass = {astro-ph},
  pages = {306--312},
  issn = {00046280, 15383873},
  doi = {10.1086/670067},
  urldate = {2025-08-26},
  archiveprefix = {arXiv}
}

@misc{greenbergAutomaticPosteriorTransformation2019,
  title = {Automatic {{Posterior Transformation}} for {{Likelihood-Free Inference}}},
  author = {Greenberg, David S. and Nonnenmacher, Marcel and Macke, Jakob H.},
  year = {2019},
  month = may,
  number = {arXiv:1905.07488},
  eprint = {1905.07488},
  primaryclass = {cs},
  publisher = {arXiv},
  doi = {10.48550/arXiv.1905.07488},
  urldate = {2025-08-24},
  archiveprefix = {arXiv}
}

@article{johnsonPredictionsNancyGrace2020,
  title = {Predictions of the {{Nancy Grace Roman Space Telescope Galactic Exoplanet Survey II}}: {{Free-Floating Planet Detection Rates}}},
  shorttitle = {Predictions of the {{Nancy Grace Roman Space Telescope Galactic Exoplanet Survey II}}},
  author = {Johnson, Samson A. and Penny, Matthew T. and Gaudi, B. Scott and Kerins, Eamonn and Rattenbury, Nicholas J. and Robin, Annie C. and Novati, Sebastiano Calchi and Henderson, Calen B.},
  year = {2020},
  month = aug,
  journal = {The Astronomical Journal},
  volume = {160},
  number = {3},
  eprint = {2006.10760},
  primaryclass = {astro-ph},
  pages = {123},
  issn = {1538-3881},
  doi = {10.3847/1538-3881/aba75b},
  urldate = {2023-02-23},
  archiveprefix = {arXiv}
}

@misc{kingmaAdamMethodStochastic2017,
  title = {Adam: {{A Method}} for {{Stochastic Optimization}}},
  shorttitle = {Adam},
  author = {Kingma, Diederik P. and Ba, Jimmy},
  year = {2017},
  month = jan,
  number = {arXiv:1412.6980},
  eprint = {1412.6980},
  primaryclass = {cs},
  publisher = {arXiv},
  doi = {10.48550/arXiv.1412.6980},
  urldate = {2025-08-24},
  archiveprefix = {arXiv}
}

@misc{lemosSamplingBasedAccuracyTesting2023,
  title = {Sampling-{{Based Accuracy Testing}} of {{Posterior Estimators}} for {{General Inference}}},
  author = {Lemos, Pablo and Coogan, Adam and Hezaveh, Yashar and {Perreault-Levasseur}, Laurence},
  year = {2023},
  month = jun,
  number = {arXiv:2302.03026},
  eprint = {2302.03026},
  primaryclass = {stat},
  publisher = {arXiv},
  doi = {10.48550/arXiv.2302.03026},
  urldate = {2025-08-24},
  archiveprefix = {arXiv}
}

@misc{mrozTESSFreefloatingPlanet2024,
  title = {{{TESS Free-floating Planet Candidate Is Likely}} a {{Stellar Flare}}},
  author = {Mroz, Przemek},
  year = {2024},
  month = may,
  number = {arXiv:2404.16480},
  eprint = {2404.16480},
  publisher = {arXiv},
  doi = {10.48550/arXiv.2404.16480},
  urldate = {2024-10-31},
  archiveprefix = {arXiv}
}

@article{pennyPredictionsWFIRSTMicrolensing2019,
  title = {Predictions of the {{WFIRST Microlensing Survey I}}: {{Bound Planet Detection Rates}}},
  shorttitle = {Predictions of the {{WFIRST Microlensing Survey I}}},
  author = {Penny, Matthew T. and Gaudi, B. Scott and Kerins, Eamonn and Rattenbury, Nicholas J. and Mao, Shude and Robin, Annie C. and Novati, Sebastiano Calchi},
  year = {2019},
  month = feb,
  journal = {The Astrophysical Journal Supplement Series},
  volume = {241},
  number = {1},
  eprint = {1808.02490},
  primaryclass = {astro-ph},
  pages = {3},
  issn = {1538-4365},
  doi = {10.3847/1538-4365/aafb69},
  urldate = {2023-10-30},
  archiveprefix = {arXiv}
}

@article{perkinsDisentanglingBlackHole2024,
  title = {Disentangling the {{Black Hole Mass Spectrum}} with {{Photometric Microlensing Surveys}}},
  author = {Perkins, Scott Ellis and McGill, Peter and Dawson, William and Abrams, Natasha S. and Lam, Casey Y. and Ho, Ming-Feng and Lu, Jessica R. and Bird, Simeon and Pruett, Kerianne and Golovich, Nathan and Chapline, George},
  year = {2024},
  month = feb,
  journal = {The Astrophysical Journal},
  volume = {961},
  number = {2},
  eprint = {2310.03943},
  primaryclass = {astro-ph},
  pages = {179},
  issn = {0004-637X, 1538-4357},
  doi = {10.3847/1538-4357/ad09bf},
  urldate = {2024-10-09},
  archiveprefix = {arXiv}
}

@misc{sumiFreeFloatingPlanetMass2023,
  title = {Free-{{Floating}} Planet {{Mass Function}} from {{MOA-II}} 9-Year Survey towards the {{Galactic Bulge}}},
  author = {Sumi, Takahiro and {koshimoto}, Naoki and Bennett, David P. and Rattenbury, Nicholas J. and Abe, Fumio and Barry, Richard and Bhattacharya, Aparna and Bond, Ian A. and Fujii, Hirosane and Fukui, Akihiko and Hamada, Ryusei and Hirao, Yuki and Silva, Stela Ishitani and Itow, Yoshitaka and Kirikawa, Rintaro and Kondo, Iona and Matsubara, Yutaka and Miyazaki, Shota and Muraki, Yasushi and Olmschenk, Greg and Ranc, Clement and Satoh, Yuki and Suzuki, Daisuke and Tomoyoshi, Mio and Tristram, Paul J. and Vandorou, Aikaterini and Yama, Hibiki and Yamashita, Kansuke},
  year = {2023},
  month = jul,
  number = {arXiv:2303.08280},
  eprint = {2303.08280},
  publisher = {arXiv},
  doi = {10.48550/arXiv.2303.08280},
  urldate = {2024-10-30},
  archiveprefix = {arXiv}
}

@misc{vaswaniAttentionAllYou2023,
  title = {Attention {{Is All You Need}}},
  author = {Vaswani, Ashish and Shazeer, Noam and Parmar, Niki and Uszkoreit, Jakob and Jones, Llion and Gomez, Aidan N. and Kaiser, Lukasz and Polosukhin, Illia},
  year = {2023},
  month = aug,
  number = {arXiv:1706.03762},
  eprint = {1706.03762},
  primaryclass = {cs},
  publisher = {arXiv},
  doi = {10.48550/arXiv.1706.03762},
  urldate = {2025-08-24},
  archiveprefix = {arXiv}
}

@article{zhangRealTimeLikelihoodFreeInference2021,
  title = {Real-{{Time Likelihood-Free Inference}} of {{Roman Binary Microlensing Events}} with {{Amortized Neural Posterior Estimation}}},
  author = {Zhang, Keming and Bloom, Joshua S. and Gaudi, B. Scott and Lanusse, Francois and Lam, Casey and Lu, Jessica R.},
  year = {2021},
  month = jun,
  journal = {The Astronomical Journal},
  volume = {161},
  number = {6},
  eprint = {2102.05673},
  primaryclass = {astro-ph, physics:physics},
  pages = {262},
  issn = {0004-6256, 1538-3881},
  doi = {10.3847/1538-3881/abf42e},
  urldate = {2024-09-23},
  archiveprefix = {arXiv}
}

@misc{zwartOriginFreefloatingObjects2024,
  title = {The Origin of Free-Floating Objects in the {{Galaxy}}},
  author = {Zwart, Simon F. Portegies},
  year = {2024},
  month = sep,
  number = {arXiv:2409.19072},
  eprint = {2409.19072},
  primaryclass = {astro-ph},
  publisher = {arXiv},
  doi = {10.48550/arXiv.2409.19072},
  urldate = {2024-10-01},
  archiveprefix = {arXiv}
}

\appendix

\section{Validation on Toy Model with Analytic Bayes Factors}
\label{sec:toy}

Before applying Evidence Networks to microlensing data, we validate the methodology on a simplified problem where ground-truth Bayes factors can be computed analytically. This toy model demonstrates that transformer embedding networks trained on binary labels alone can recover calibrated Bayes factor predictions for low SNR bump-hunt timeseries problems.

\subsection{Toy Model Setup}

\begin{figure*}[!t]
    \centering
    \includegraphics[width=0.95\linewidth]{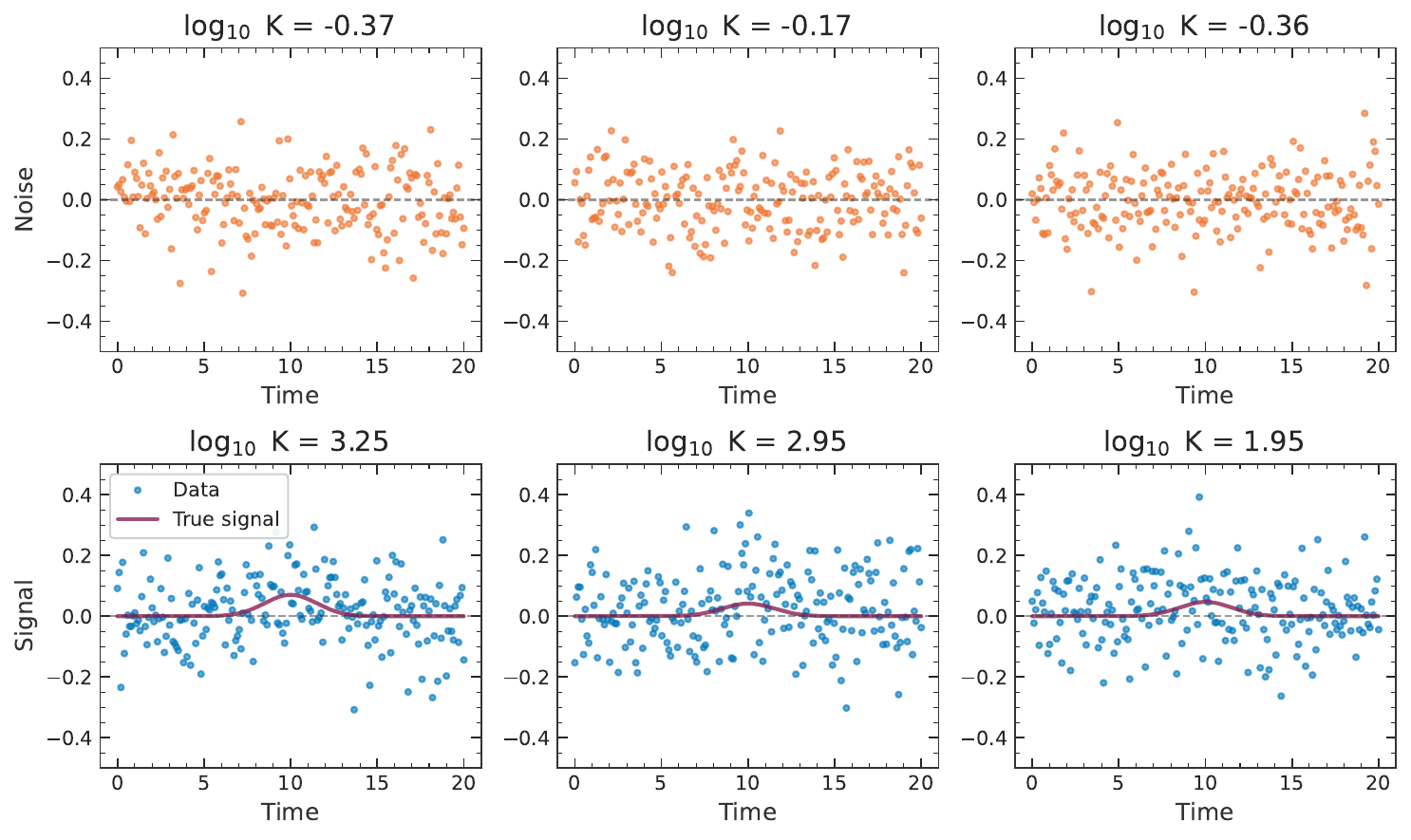}
    \caption{Toy Data Examples. The top row are noise-only and the bottom row are high Bayes-factor signals with the true underlying signal shown in red.}
    \label{fig:toy_examples}
\end{figure*}

We construct a signal detection problem using Gaussian bump templates embedded in white noise. The observed time series $\mathbf{d} \in \mathbb{R}^T$ follows one of two models:

\begin{itemize}
    \item \textbf{Signal hypothesis} $\mathcal{M}_1$: $\mathbf{d} = A \mathbf{g} + \mathbf{n}$, where $A \sim \mathcal{N}(\mu_A, \sigma_A^2)$ is a random amplitude, $\mathbf{g}$ is a unit-norm Gaussian template, and $\mathbf{n} \sim \mathcal{N}(\mathbf{0}, \sigma_n^2 \mathbf{I})$ is white noise
    \item \textbf{Noise-only hypothesis} $\mathcal{M}_0$: $\mathbf{d} = \mathbf{n}$
\end{itemize}

For this linear-Gaussian model, the Bayes factor admits a closed-form solution. Defining the matched-filter statistic $x \equiv \mathbf{d}^\top \mathbf{g}$, the template norm $G \equiv \mathbf{g}^\top \mathbf{g}$, and the parameter $\alpha \equiv \sigma_A^2 G / \sigma_n^2$, the natural logarithm of the Bayes factor is:

\begin{equation}
\log K = -\frac{1}{2} \log(1 + \alpha) + \frac{1}{2} \frac{\alpha}{1 + \alpha} \rho^2
\end{equation}

\noindent where $\rho^2 \equiv x^2 / (\sigma_n^2 G)$ is the matched-filter signal-to-noise ratio squared. This analytic expression serves as ground truth for validation.

We generate toy data with $T = 200$ time points over a 20-day window with noise standard deviation $\sigma_n = 0.1$.  Gaussian templates are centered and have widths of $1.5$ days, with amplitudes drawn from the prior $A \sim \mathcal{N}(0.1, 0.1^2)$.  We train the Evidence Network on 50,000 simulated light curves with equal signal/noise fractions, using binary labels only. The analytic $\log K$ values are reserved strictly for post-training evaluation.

The Evidence Network uses the same transformer encoder described in Section~\ref{sec:architecture} (6 layers, 8 heads, $d_{\text{model}} = 64$), followed by a 2-layer MLP classifier with hidden dimension 64. We train using the l-POP-Exponential loss with $\alpha = 2$ for 60 epochs, employing an Adam optimizer with learning rate $10^{-4}$ and a step learning rate scheduler (factor 0.5 every 20 epochs without improvement).

\subsection{Toy Model Performance}

\begin{figure*}[!t]
    \centering
    \begin{subfigure}[t]{0.48\textwidth}
        \centering
        \includegraphics[width=\textwidth]{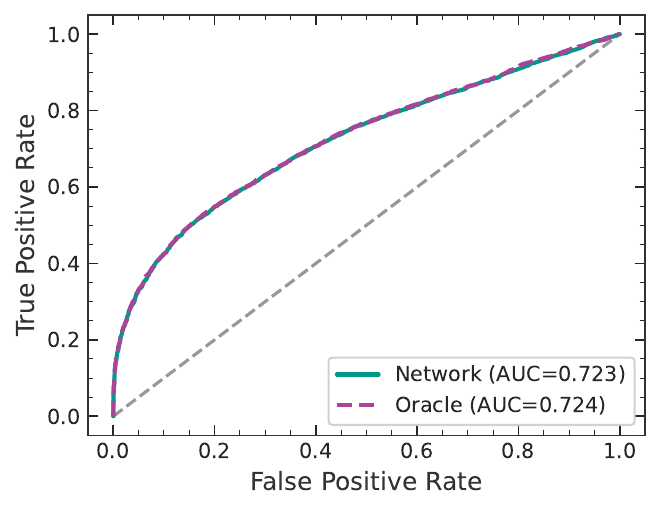}
        \caption{ROC curve comparing Evidence Network to oracle classifier}
        \label{fig:toy_roc}
    \end{subfigure}
    \hfill
    \begin{subfigure}[t]{0.48\textwidth}
        \centering
        \includegraphics[width=\textwidth]{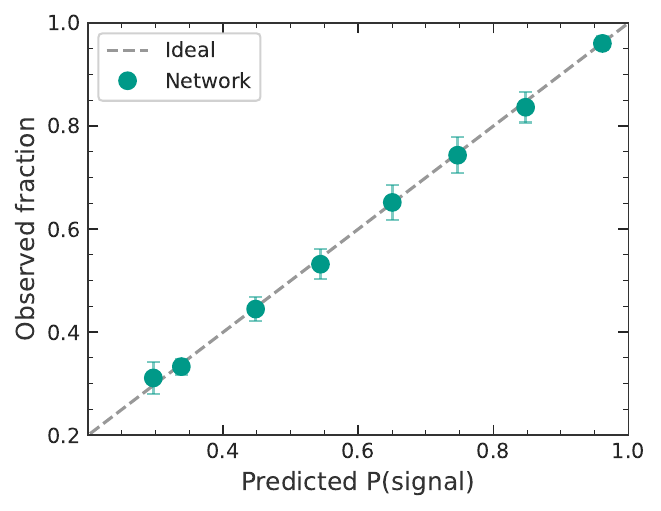}
        \caption{Calibration: predicted vs. observed signal fraction}
        \label{fig:toy_calibration}
    \end{subfigure}
    \caption{Toy model validation with analytic Bayes factors. \textbf{(a)} The Evidence Network achieves near-oracle ROC performance (AUC=0.725 vs 0.726), demonstrating it extracts essentially all discriminative information. \textbf{(b)} Calibration curve shows predicted model posteriors match empirical signal fractions, confirming the network produces well-calibrated Bayes factor estimates from binary labels alone.}
    \label{fig:toy_validation}
\end{figure*}

Figure~\ref{fig:toy_roc} shows the receiver operating characteristic (ROC) curve for the trained Evidence Network compared against an ``oracle'' classifier that uses the analytic $\log_{10} K$ directly. The network achieves near-oracle discriminative performance (AUC = 0.725 vs.\ oracle 0.726), demonstrating that Evidence Networks recover calibrated Bayes factors from binary labels alone without requiring explicit likelihood computation. The near-perfect overlap of the ROC curves shows that the network has learned to extract essentially all available statistical information for discriminating signal from noise. The network's true positive rate at 1\% false positive rate is 0.32, compared to the oracle's 0.33. At higher false positive rates (5\% and 10\%), the network achieves values within 0.01 of the oracle, confirming robust performance across the full operating range.

While ROC curves measure discriminative performance, calibration assesses whether predicted probabilities match empirical frequencies, a crucial requirement for Bayesian model comparison. We apply the coverage test of \citet{jeffreyEvidenceNetworksSimple2024}: events are binned by predicted model posterior $p(\mathcal{M}_1|\mathbf{x}) = K/(1+K)$, and well-calibrated predictions satisfy $p(\mathcal{M}_1|\mathbf{x}) \approx f_{\text{signal}}$ within binomial uncertainty $\sigma_{\text{binom}} = \sqrt{f_{\text{signal}}(1 - f_{\text{signal}})/N_{\text{bin}}}$, where $f_{\text{signal}}$ is the empirical fraction of true signals in each bin. Figure~\ref{fig:toy_calibration} shows the resulting calibration curve. The points lie almost exactly on the identity line, with deviations consistent with binomial sampling uncertainty.

These results motivate application of Evidence Networks to microlensing detection, where we similarly have access to labeled training data (from simulations) but lack analytic expressions for Bayes factors.

\section{Representative Events Detected by the Evidence Network}
\label{app:network_only}

\begin{figure*}[!t]
    \centering
    \includegraphics[width=0.95\linewidth]{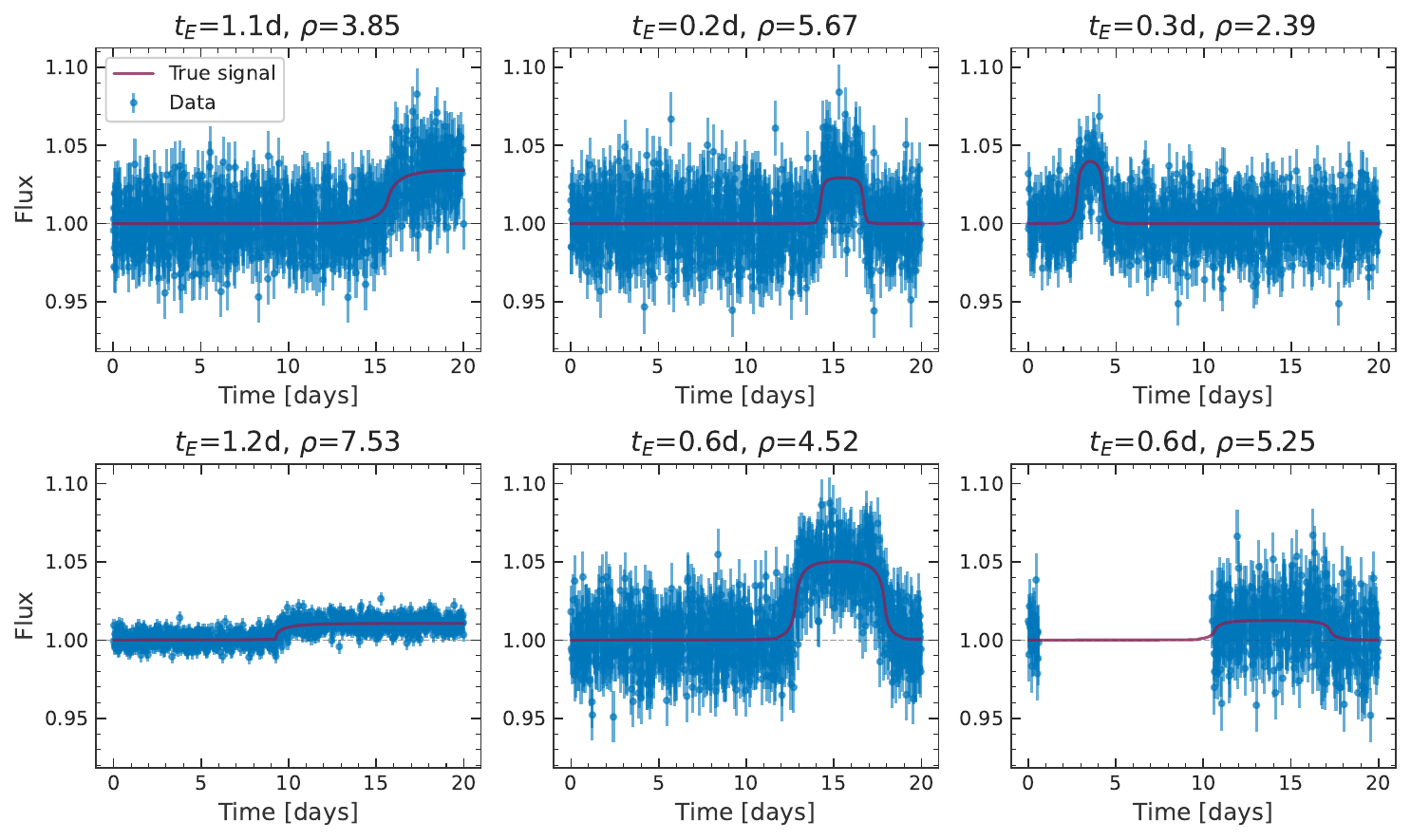}
    \caption{Representative events detected by the Evidence Network (with $\log_{10} K > 14$, strong detections) but missed by hard cuts from \citet{johnsonPredictionsNancyGrace2020}. These examples illustrate systematic failures of threshold-based detection: low peak magnification (top left, middle), observation gaps interrupting consecutive high-SNR points (top right, bottom left), and ongoing events without established baseline (bottom middle, right). The Evidence Network extracts signal from the full temporal morphology rather than requiring specific summary statistics to exceed fixed thresholds.}
    \label{fig:network_only}
\end{figure*}

To make the failure modes of hard cuts concrete, Figure~\ref{fig:network_only} shows six finite-source events that the Evidence Network assigns strong detections ($\log_{10} K > 14$) but that are missed by the \citet{johnsonPredictionsNancyGrace2020} criteria. All six lie in the short-duration, large-$\rho$ regime ($t_E \lesssim 1$ day, $\rho \gtrsim 2$) that dominates the low-mass FFP population.

Three distinct failure modes appear. First, finite-source broadening spreads the magnification signal over several days, so even with clear morphological structure the peak rarely clears the $3\sigma$ threshold needed for $N_{\text{consec}} \geq 6$ consecutive points (top left, top middle). Second, observation gaps coincident with the peak interrupt the run of high-SNR points, so a detectable event fails the consecutivity requirement despite an unambiguous signal in the ungapped data (top right, bottom left). Third, events whose peak falls near the edge of the observing window never establish a clean pre- or post-event baseline, causing the $\Delta\chi^2$ reference fit to be biased and the cut to reject the event (bottom middle, bottom right).

In all three cases the signal remains extractable because the transformer attends to the full light curve rather than to a fixed summary statistic. This is why the Evidence Network's advantage is concentrated precisely in the $\rho \gtrsim 5$ regime most relevant to the FFP mass function.

\section{Shared- vs.\ Separate-Encoder Ablation and Re-Tuned Hard Cuts}
\label{app:ablation}

\paragraph{Does sharing the encoder help?} To test whether the shared transformer encoder is beneficial or merely architectural, we train an Evidence detection head on top of the \emph{frozen} encoder learned by the NPE (inference) model, and compare it head-to-head against a dedicated Evidence model trained from scratch. Both are evaluated on the same seeded test set of 5,000 signals and 5,000 Gaussian-noise events, with the detection threshold on $\log_{10}K$ chosen per model so that the validation false-positive rate is zero. We run the comparison at two training budgets: 20K-samples and 100K-samples. Table~\ref{tab:ablation} reports the result.

\begin{table}[H]
\centering
\caption{Transfer vs. dedicated encoder detection performance at two training budgets. At both budgets the transfer model reaches $99.9\%$; the from-scratch detector matches it only approximately, and only at the larger budget.}
\label{tab:ablation}
\small
\begin{tabular}{@{}llrr@{}}
\toprule
Budget & Model & $\log_{10}K$ thr. & TPR \\
\midrule
20K  & Dedicated & 18.2 & $93.0\%$ \\
20K  & Transfer  & 1.20 & $99.9\%$ \\
100K & Dedicated & 1.80 & $98.9\%$ \\
100K & Transfer  & 0.11 & $99.9\%$ \\
\bottomrule
\end{tabular}
\end{table}

Two readings follow. First, at the small 20K budget the transfer model already matches the headline $99.9\%$ detection rate while the matched from-scratch detector lags at $93.0\%$: the SBI-trained embedding does the heavy lifting and the head needs only a small classification budget. Second, at the matched 100K budget the from-scratch detector closes most of the gap ($98.9\%$ vs.\ $99.9\%$), but the transfer model still wins. The transfer model also reaches FPR $=0$ at a smaller threshold; its noise $\log_{10}K$ distribution is consistent with the inference-trained encoder having absorbed information about the full microlensing manifold that a classification-only loss does not capture. Finally, on identical hardware and hyperparameters, the transfer model trains $3.8\times$ faster than the from-scratch model.

\paragraph{Re-tuned hard cuts.} We also re-tune each hard cut on a validation set drawn from the same generator used to train the network, jointly sweeping the statistic thresholds. Table~\ref{tab:tuned_cuts} reports both the literature and re-tuned criteria. Re-tuning raises every cut's TPR to $\sim$$97\%$ (from $91$--$94\%$), but all remain below the Evidence Network's $99.9\%$. The hard cuts retain the advantage of oracle (true) lensing parameters throughout.

\begin{table}[H]
\centering
\caption{Literature vs.\ re-tuned hard-cut detectors on the seeded test set. ``Stat'' / ``Consec'' are the statistic and consecutive-point thresholds. TPR is the detection rate; FPR is on the Gaussian-noise test set. Hard cuts use oracle parameters.}
\label{tab:tuned_cuts}
\resizebox{\columnwidth}{!}{%
\small
\begin{tabular}{@{}lcccc@{}}
\toprule
Method & Lit.\ stat / consec & Tuned stat / consec & Lit.\ TPR & Tuned TPR (FPR) \\
\midrule
\citet{johnsonPredictionsNancyGrace2020} & $\Delta\chi^2\!\geq\!300$ / 6 & $300$ / 2 & $95.9\%$ & $97.0\%$ ($0$) \\
\citet{sumiUnboundDistantPlanetary2011} & $\chi_{3+}\!\geq\!80$ / 3 & $6.98$ / 2 & $93.9\%$ & $97.2\%$ ($2\times10^{-4}$) \\
\citet{mrozNoLargePopulation2017a} & $\chi_{3+}\!\geq\!32$ / 3 & $6.98$ / 2 & $95.4\%$ & $97.2\%$ ($2\times10^{-4}$) \\
\bottomrule
\end{tabular}%
}
\end{table}

\section{NPE vs.\ MCMC Wall-Clock Benchmark}
\label{app:npe_timing}

We measure wall-clock cost directly. We compare amortized NPE sampling against an \texttt{emcee} \citep{foreman-mackeyEmceeMCMCHammer2013} sampler on a single NVIDIA H100. The MCMC budget is $5{,}000$ retained samples per event (32 walkers, $1{,}000$-step burn-in). Table~\ref{tab:npe_timing} reports per-event timings. The median speedup is ${\sim}16{,}000\times$.

\begin{table}[H]
\centering
\caption{Per-event wall-clock cost of NPE sampling vs.\ \texttt{emcee} MCMC on the same forward model (NVIDIA H100, 20 events, $10{,}000$ NPE samples / $5{,}000$ MCMC samples per event).}
\label{tab:npe_timing}
\small
\begin{tabular}{@{}lrr@{}}
\toprule
Quantity & Median & Mean \\
\midrule
NPE per event   & $21$~ms  & $44$~ms \\
MCMC per event  & $346$~s  & $333$~s \\
Speedup (MCMC/NPE) & $16{,}400\times$ & $14{,}500\times$ \\
\bottomrule
\end{tabular}
\end{table}

We report wall-clock cost only. The MCMC time is for single-event sampling with a modest walker count and a CPU-side likelihood evaluator. Optimizing the MCMC configuration could shrink this gap, but amortized inference will continue to be much faster.

\section{Posterior Bias Decomposition}
\label{app:bias}

Table~\ref{tab:posterior_metrics} summarizes the per-parameter posterior accuracy and calibration over all 5,000 test events. The well-constrained parameters ($t_0$, $\log_{10}t_E$) have small bias; the degeneracy-affected parameters ($u_0$, $\log_{10}\rho$, $f_s$) carry larger residuals, with $95\%$ credible-interval coverage remaining between $0.90$ and $0.97$ throughout.

\begin{table}[H]
\centering
\caption{Per-parameter posterior diagnostics over 5,000 test events: median bias, RMSE, Pearson correlation between injected and recovered medians, and empirical coverage of the $68\%$ and $95\%$ credible intervals.}
\label{tab:posterior_metrics}
\small
\begin{tabular}{@{}lrrrrr@{}}
\toprule
Parameter & Bias & RMSE & $r$ & 68\% & 95\% \\
\midrule
$t_0$            & $+0.010$  & $0.445$ & $0.997$ & $0.826$ & $0.972$ \\
$u_0$            & $+0.084$  & $0.222$ & $0.770$ & $0.636$ & $0.925$ \\
$\log_{10}t_E$   & $-0.018$  & $0.064$ & $0.994$ & $0.661$ & $0.931$ \\
$\log_{10}\rho$  & $+0.056$  & $0.436$ & $0.824$ & $0.553$ & $0.901$ \\
$f_s$            & $+0.029$  & $0.131$ & $0.881$ & $0.643$ & $0.916$ \\
\bottomrule
\end{tabular}
\end{table}

\begin{figure}[H]
    \centering
    \includegraphics[width=\linewidth]{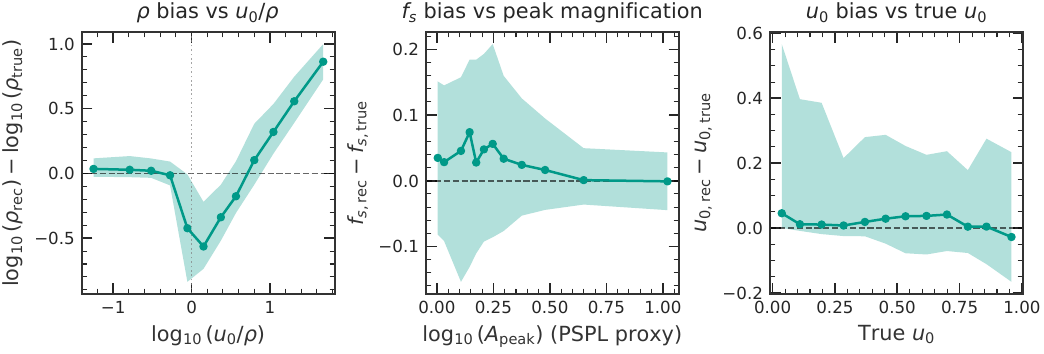}
    \caption{Posterior bias decomposition. \textbf{Left:} $\log_{10}\rho$ residual vs.\ $u_0/\rho$. \textbf{Middle:} $f_s$ residual vs.\ peak magnification. \textbf{Right:} $u_0$ residual vs.\ true $u_0$. Bands show $16$--$84$th percentile residuals per bin.}
    \label{fig:bias}
\end{figure}

Figure~\ref{fig:bias} decomposes these residuals against the physically relevant axes. The $\log_{10}\rho$ bias is consistent with zero in the well-resolved finite-source regime ($u_0 \gg \rho$) and develops a structured bias only as $u_0 \sim \rho$, exactly where the $u_0\!\leftrightarrow\!\rho$ degeneracy folds two physical solutions onto one light curve. The $f_s$ bias shrinks with peak magnification, as expected from the $(f_s, A_{\text{peak}})$ degeneracy. Finally, $u_0$ recovery is essentially unbiased for $u_0 > 0.2$ and, near the detection boundary ($u_0 \to 1$, $u_0 \sim \rho$), inflates its credible interval rather than producing a point bias. The network widens its posterior where the data are uninformative rather than becoming overconfident. 

\end{document}